\documentstyle[12pt]{article}
\def\inbar{\vrule height1.5ex width.4pt depth0pt}
\def\IB{\relax{\rm I\kern-.18em B}}
\def\IC{\relax\,\hbox{${\rm I}\kern-.5em{\rm C}$}}
\def\ID{\relax{\rm I\kern-.18em D}}
\def\IE{\relax{\rm I\kern-.18em E}}
\def\IF{\relax{\rm I\kern-.18em F}}
\def\IG{\relax\,\hbox{$\inbar\kern-.3em{\rm G}$}}
\def\IH{\relax{\rm I\kern-.18em H}}
\def\II{\relax{\rm I\kern-.18em I}}
\def\IK{\relax{\rm I\kern-.18em K}}
\def\IL{\relax{\rm I\kern-.18em L}}
\def\IM{\relax{\rm I\kern-.18em M}}
\def\IN{\relax{\rm I\kern-.18em N}}
\def\IO{\relax\,\hbox{$\inbar\kern-.3em{\rm O}$}}
\def\IP{\relax{\rm I\kern-.18em P}}
\def\IQ{\relax\,\hbox{$\inbar\kern-.3em{\rm Q}$}}
\def\IR{\relax{\rm I\kern-.18em R}}
\def\ZZ{\relax{\sf Z\kern-.4em Z}}
\def\fnote#1#2{\begingroup\def\thefootnote{#1}\footnote{#2}\addtocounter
{footnote}{-1}\endgroup}
\def\beq{\begin{equation}}
\def\eeq{\end{equation}}
\def\bea{\begin{eqnarray}}
\def\eea{\end{eqnarray}}
\def\lleq#1{\label{#1}\eeq}
\def\llea#1{\label{#1}\eea}
\let\nn=\nonumber
\def\tabroom{\hbox to0pt{\phantom{\Huge A}\hss}}
\def\notin{\ \hbox{{$\in$}\kern-.51em\hbox{/}}}
\def\a{\alpha}   \def\b{\beta}     
     \def\l{\lambda}

     \def\si{\sigma}
   \def\th{\theta}

\def\cN{{\cal N}}

\def\tchi{\tilde \chi}

   \def\bz{{\bar z}}

 \def\bPhi{{\bar \Phi}}

\def\bth{{\bar \theta}}

\def\vcrit{V_{crit}} \def\mvcrit{\Lambda_{crit}}

\def\b2{h^{(1,1)}}
\def\b3{h^{(2,1)}}
\def\b4{h^{(2,2)}}
\def\b5{h^{(3,2)}}
\def\b6{h^{(4,1)}}

\begin{document}

\hfill{BONN--HE--93--47}
\vskip .01truein
\hfill{NSF--ITP--93--146}

\vskip 1.2truein

\centerline {\large K\"AHLER MANIFOLDS WITH POSITIVE FIRST CHERN CLASS}
\vskip .1truein
\centerline{\large  AND MIRRORS OF RIGID CALABI--YAU MANIFOLDS
            \fnote{\diamond}{Based on a talk presented at the International
                             Workshop on Supersymmetry and Unification of
                             Fundamental Interactions (SUSY 93),
                             Boston, MA, March 1993.}
            \fnote{\star}{This work was supported in part by NSF grant
                          PHY--89--04035}
           }

\vskip .9truein
\centerline{\sc Rolf Schimmrigk
                \fnote{**}{Email address: netah@avzw02.physik.uni-bonn.de}
           }
\vskip .1truein
\centerline{\it Physikalisches Institut, Universit\"at Bonn, Nussallee 12}
\vskip .05truein
\centerline{\it 53115 Bonn 1, Germany}

\vskip 1.6truein
\centerline{\bf Abstract}

 Recent work on the relation between a special class of
 K\"ahler manifolds with positive first Chern class and
 critical N$=$2 string vacua with c$=$9 is reviewed and
 extended.

\renewcommand\thepage{}
\vfill
\eject

\parindent=20pt
\baselineskip=18pt
\pagenumbering{arabic}
\parskip=10pt

\noindent
{\bf  1. Introduction}

\noindent
To answer the question whether mirror symmetry [1][2][3] is in fact a
symmetry of string theory is nontrivial for two rather different
reasons. In the absence of a universal mirror construction there are
`technical' difficulties, related to the explicit construction of
mirror pairs. More importantly, there might exist a
`fundamental obstruction'
which forbids the existence of mirrors for particular vacua from
first principles and hence might lead to a restriction of the class
 of string vacua for which mirror symmetry is an allowed operation.

At first sight it appears that such a fundamental obstruction is
furnished by a small, but prominent, class of vacua: toroidal orbifolds.
Such manifolds lead to so--called rigid vacua, i.e. groundstates which
 do not have complex deformations. This means the following:
Recall that because Calabi--Yau spaces are complex manifolds they admit a
 Hodge decomposition of their real de Rham cohomology groups. Therefore
the Betti numbers
$b_i={\rm dim}_{\IR}~ {\rm H_{DR}}^i(M,\IR)$, $i=0,1,...,dim_{\IR} M$,
 split into the
Hodge numbers $h^{(p,q)}={\rm dim}_{\IC}~ {\rm H}^{p,q}(M, \IC)$,
$p,q=0,...,dim_{\IC} M$:
\beq
b_i(M) = \sum_{p+q=i}
            h^{(p,q)}(M).
\eeq
Because Calabi--Yau spaces are K\"ahler the Hodge numbers are symmetric,
$h^{(p,q)}=h^{(q,p)}$, and because the first Chern class vanishes
it follows
that $h^{(p,0)}=0=h^{(0,p)}$ for  $p=1,2$ and $h^{(3,0)}=1=h^{(0,3)}$.
Hence the cohomology of the internal space
consists of only two independent groups, the K\"ahler sector
 ${\rm H}^{(1,1)}\equiv {\rm H}^{(2,2)}$ and the complex deformation sector
${\rm H}^{(2,1)} \equiv {\rm H}^{(1,2)}$ which parametrize the number
of antigenerations
and generations, respectively, that are observed in low energy physics.

It has been observed in [1][2] in the framework of Landau--Ginzburg
vacua and in  [3] in the class of Gepner models, that the space
of (2,2)--vacua features mirror symmetry, i.e. for critical
string vacua $\vcrit$ with Hodge numbers
\beq
\vcrit: (\chi, h^{(1,1)}, h^{(2,1)})
\eeq
there exists a mirror partner in which the Hodge numbers are
exchanged
\beq
\mvcrit: (-\chi, h^{(2,1)}, h^{(1,1)}),
\eeq
defining what is called a mirror spectrum of $V_{crit}$.
An explicit construction relating different Landau--Ginzburg theories
via a path integral argument that involves fractional transformations
of the order parameters has been provided in [2].

Rigid Calabi--Yau vacua without (2,1)--cohomology then appear to furnish
a fundamental obstruction to mirror symmetry:
Since the mirror map exchanges complex deformations and
K\"ahler deformations of a manifold
it would seem that the mirror of a rigid Calabi--Yau manifold cannot be
K\"ahler and hence does not describe a consistent string vacuum. In
fact, it appears, using Zumino's result [4] that $N=2$
supersymmetry of a $\si$--model requires the target manifold to be
 K\"ahler, that the mirror vacuum cannot even be $N=1$ spacetime
supersymmetric. It follows that the class of Calabi--Yau manifolds is
not the appropriate setting for mirror symmetry and
the question arises  what the proper framework might be.

This review describes and extends recent work [5] which
shows the existence of a new class of manifolds
which generalizes the class of Calabi--Yau spaces of complex dimension
$D_{crit}$ in a natural way. The manifolds involved are of complex
dimension $(D_{crit}+2(Q-1))$, $Q\in \IN$ and have a positive
first Chern class which is quantized in
multiples of the degree of the manifold. Thus they do not describe,
a priori, consistent string groundstates. It was shown in [5]
however, that it is
possible to derive from these higher dimensional
manifolds the spectrum of  critical string vacua. This
can be done not only for the generations but also for the
antigenerations. For particular types of these new manifolds it is
also possible to construct the corresponding
$D_{crit}$--dimensional Calabi--Yau manifold directly from the
$(D_{crit}+2(Q-1))$--dimensional space.

It should be emphasized that the noncritical manifolds described
below do not just provide an alternative realization of the
physical modes observed in four dimensions: even though the
spectrum of the critical vacuum $V_{crit}$, parametrized by
(generalized, if necessary) cohomology groups, is embedded in
the Hodge diamond of the noncritical manifold
$M_{D_{crit}+2(Q-1)}$
\beq
H^{(p,q)}(V_{D_{crit}}) \subset H^{(p+Q-1,q+Q-1)}(M_{D_{crit}+2(Q-1)}),
\eeq
the cohomology groups of the higher dimensional manifold are
generically larger than those of the critical groups and hence
they contain additional modes,
which at present do not have a physical interpretation.
The main result of [5] is a geometrical construction which
projects out the critical modes.

An important feature of the higher dimensional cohomology is a sort
of `doubling phenomenon': It can happen that more than one copy of
the complete critical cohomology is embedded in the Hodge diamond
of the noncritical manifold. In particular the  Hodge--diamond for
noncritical manifolds of odd complex dimension will, in general,
lead to a middle Betti number of the form
\beq
b_{D_{crit}+2(Q-1)}=2n+2\sum_{i=1}^{D_{crit}-1}
h^{(D_{crit}+(Q-1)-i,(Q-1)+i)}
\eeq
where $n$ is some integer larger than zero which can be larger than
unity. The reason why this can happen is explained by the
construction introduced in [5], as will become clear in
Section 4.

This new  class of manifolds is not in one to one correspondence
with the class of Calabi--Yau manifolds as it also contains manifolds
which describe string vacua
which do not contain massless modes corresponding to antigenerations.
It is precisely this new type of manifold that is
needed in order to construct mirrors of rigid Calabi--Yau manifolds
 without generations.

\vskip .3truein
\noindent
{\bf 2. Higher Dimensional Manifolds with Quantized
Positive First Chern Class }

\noindent
The construction of  noncritical manifolds proceeds via the following
prescription [5]:
\begin{itemize}
\item Fix the central charge $c$ of the (2,2)--vacuum states
     and its critical dimension
 \beq
  D_{crit} = c/3.
 \eeq
\item Choose a positive integer $Q\in \IN$.
\item Introduce $(D_{crit}+2Q)$ complex coordinates
     $(z_1,...,z_{D_{crit}+2Q}), z_i \in \IC$.
\item Define an equivalence relation
   \beq
     (z_1,...,z_{D_{crit}+2Q})
     \sim (\l^{k_1}z_1,...,\l^{D_{crit}+2Q} z_{D_{crit}+2Q})
   \eeq
 where $\l \in \IC^*$ is a nonzero complex number and the
positive integers  $k_i \in \IN$ are the weights of these coordinates.
The set of these
equivalence classes defines so--called weighted projective spaces,
compact manifolds which are denoted by $\IP_{(k_1,...,k_{D_{crit}+2Q})}$.
\item Define hypersurfaces in the ambient weighted projective space
    by imposing a constraint defined by polynomials $p$ of degree
  \beq
   d= \frac{1}{Q} \sum_i k_i
  \lleq{pquant}
  i.e. $p(\l^i z_i) = \l^d p(z_i)$.
\end{itemize}

\noindent
The family of hypersurfaces embedded in the ambient space
 as the zero locus of $p$ will be denoted by
 \bea
   M_{D_{crit}+2(Q-1)} &=&  \{p(z_1,\dots,z_{D_{crit}+2Q})=0\}~\cap ~
                         \IP_{(k_1,\dots, k_{D_{crit}+2Q})} \nn \\
                   &=&  \IP_{(k_1,...,k_{D_{crit}+2Q})}
                      \left[ \frac{1}{Q} \sum_{i=1}^{D_{crit}+2Q} k_i\right]
\llea{newmfs}
and is  called a configuration.

The constraint (\ref{pquant}) is the essential defined property, which
links the degree of the polynomial to the weights of the ambient space
and  is a rather
restrictive condition in that it excludes many types of varieties which
are transverse but are not of physical relevance
\footnote{A subclass of the manifolds described by (\ref{newmfs}) has
        recently also been discussed in [6] and a
        particular simple manifold in this class, the cubic sevenfold
       $\IP_8[3]$, has been analyzed in detail in [6][7][8].}.
A simple example is the Brieskorn--Pham type hypersurface
\beq
\IP_{(420,280,210,168,140,120,105)}[840] \ni
\left\{p= \sum_{i=1}^7 z_i^{i+1} =0 \right\}
\lleq{excluded}
which is a transverse, i.e. quasismooth manifold. It is
also interesting from a different point of view: An important feature of
Calabi--Yau hypersurfaces is that they are automatically what is called
{\it well formed}, i.e. they do not contain orbifold singularities that
are surfaces (in the case of threefolds). More generally this fact
translates into the statement that the only resolutions that have to
be performed are so--called small resolutions, i.e. the singular sets
are of codimension larger than one. The same is true for the
higher dimensional manifolds defined above whereas the manifold
(\ref{excluded})  contains the
singular 4--fold $S=\IP_{(210,140,105,84,70,60)}[420]$.

Alternatively, manifolds of the type above may be characterized
via a curvature constraint. Because of (\ref{pquant}) the first
Chern class is given by
\beq
c_1(M_{D_{crit}+2(Q-1)}) =(Q-1)~c_1(\cN)
\lleq{c1quant}
where $c_1(\cN)=dh$ is the first Chern class of the normal bundle
$\cN$ of the hypersurface $M_{D_{crit}+2(Q-1)}$ and $h$ is the
pullback of the K\"ahler form
${\rm H}\in {\rm  H}^{(1,1)}
\left(\IP_{(k_1,\dots,k_{D_{crit}+2Q})}\right)$
of the ambient space. Hence the first Chern class is quantized in
multiples of the degree of the hypersurface $M_{D_{crit}+2(Q-1)}$.
For $Q=1$ the first Chern class vanishes and the manifolds
for which (\ref{pquant}) holds are Calabi--Yau manifolds, defining
consistent groundstates of the supersymmetric closed string.
For $Q > 1$ the first Chern class is nonvanishing and therefore these
manifolds cannot possibly describe vacua of the critical string, or
so it seems.

 It has been shown in [5] however that it is possible to
derive from these higher dimensional manifolds the massless spectrum
of critical vacua. It is furthermore possible, for certain subclasses
of hypersurfaces of type (\ref{newmfs}), to construct
Calabi--Yau manifolds $M_{CY}$ of dimension $D_{crit}$  and complex
codimension
\beq
codim_{\IC} (M_{CY}) =Q      \nn
\eeq
directly from these manifolds.
The integer $Q$ thus plays a central r\^{o}le:
the critical dimension is the dimension of the noncritical manifolds offset
by twice the coefficient of the
first Chern class of the normal bundle of the hypersurface, which
involves $Q$. The physical interpretation of the integer $Q$ is that
of a total charge associated to the corresponding Landau--Ginzburg
theory which determines the codimension of the Calabi--Yau manifold
which it describes.

As mentioned in the introduction the class of spaces defined by
(\ref{newmfs}) contains manifolds with no antigenerations
and hence it is necessary to have some way other than Calabi--Yau
manifolds to represent string groundstates if one wants to compare
them with the higher
dimensional manifolds. One possible way to do this is to relate them to
Landau--Ginzburg theories:  manifolds of type (\ref{newmfs}) can be
viewed as a projectivization via a weighted equivalence defined on an
affine noncompact hypersurface  defined by the same polynomial
\beq
\IC_{(k_1,...,k_{D_{crit}+2Q})}\left[d\right] \ni
\{p(z_1,...,z_{D_{crit}+2Q})=0\}.
\lleq{affvar}
Because the polynomial $p$ is assumed to be transverse in the projective
ambient
space the affine variety has a very mild singularity: it has an isolated
singularity at the origin defining what is called a catastrophe in the
mathematics literature.

The complex variables $z_i$ parametrizing the ambient space are to be
viewed as the field theoretic limit
$\varphi_i(z,\bz) = z_i$ of the lowest components of the order
parameters
$\Phi_i(z_i,\bz_i,\th^{\pm}_i,\bth^{\pm}_i)$,
described by chiral $N=2$ superfields of a 2--dimensional
Landau--Ginzburg theory defined by the action
\beq
\int d^2z d^2\th d^2\bth~ K(\Phi_i,\bPhi_i) +
\int d^2z d^2\th~ W(\Phi_i) + c.c.                      \nn
\eeq
where $K$ is the K\"ahler potential and $W$ is the superpotential.
It was shown in [9][10] that such Landau--Ginzburg
theories are useful for the understanding of string
vacua and also that much information about such groundstates is already
encoded in the associated catastrophe (\ref{affvar}). A crucial
piece of
information about a vacuum, e.g., is its central charge. Using a result
from singularity theory, it is easy to derive that the central charge
of the conformal fixed point of the LG theory is
\beq
c=3\sum_{i=1}^{D_{crit}+2Q} \left(1-2q_i\right),
\eeq
where $q_i =k_i/d$ are the U(1) charges of the superfields. It is
furthermore  possible
to derive the massless spectrum of the GSO projected fixed of the
LG theory, defining
the string vacuum, directly from the catastrophe (\ref{affvar}) via a
procedure described by Vafa [11].

Even though the manifolds (\ref{newmfs}) therefore correspond to
LG theories of central charge $c=3D_{crit}$ they can, however,
not be identical to such theories: Consider the
case when the critical dimension of the internal space corresponds
to our world, i.e. $D_{crit}=3$ and $Q=2$. The cohomology
of $M_5$ leads to the Hodge diamond

\begin{small}
\begin{center}
\begin{tabular}{c c c c c c c c c c c}
  &  &      &   &          &1         &         &   &       &   &  \\
  &  &      &   &0         &          &0        &   &       &   &  \\
  &  &      &0  &          &$(1,1)$     &         &0  &       &   &  \\
  &  &0     &   &0         &          &0        &   &0      &   &  \\
  &0 &      &0  &          &$(2,2)$     &         &0  &       &0  &  \\
0 &  &$(4,1)$ &   &$(3,2)$     &          &$(3,2)$    &   &$(4,1)$  &   &0 \\
  &0 &      &0  &          &$(2,2)$     &         &0  &       &0  &  \\
  &  &0     &   &0         &          &0        &   &0      &   &  \\
  &  &      &0  &          &$(1,1)$     &         &0  &       &   &  \\
  &  &      &   &0         &          &0        &   &       &   &  \\
  &  &      &   &          &1         &         &   &       &   &  \\
\end{tabular}
\end{center}
\end{small}
where $(p,q)$ denotes the dimension $h^{(p,q)}$ of the cohomology group
${\rm H}^{(p,q)}(M_5)$.

It is clear from this Hodge diamond that the higher dimensional manifolds
will contain more modes than the critical vacuum and hence the relation
of the spectrum of the critical vacuum and the cohomology of the noncritical
manifolds will be a nontrivial one.

\vskip .3truein
\noindent
{\bf 3. Noncritical Manifolds and Critical Vacua}

\noindent
In certain benign situations the subring of monomials
of charge unity in the chiral ring describes the generations of the
vacuum [12]. For this to hold at all it is important that
the GSO projection
is the canonical one with respect to the cyclic group $\ZZ_d$,
the  order of which is the degree $d$ of the superpotential
\footnote{It does not hold for projections that involve orbifolds
with respect to different groups such as those discussed in [13].
This is   to be expected as these modified projections
       can be understood as orbifolds of canonically constructed
       vacua. The additional moddings generate singularities the
resolution of which introduces, in general,
  additonal modes in both sectors, generations and antigenerations.}.
Thus the generations are easily derived for this subclass of
 theories in (\ref{newmfs})  because the polynomial ring is identical
to the chiral ring of the corresponding Landau--Ginzburg theory.
In general a more complicated analysis, involving the singularity
structure of the  higher dimensional manifolds, will have to be
done [14].

It remains to extract the second cohomology. In a Calabi--Yau manifold
there are no holomorphic  2--forms and hence all of the second
cohomology
is in ${\rm H}^{(1,1)}$. Because of Kodaira's vanishing theorem the same
is true
for manifolds with positive  first Chern class and therefore for the
manifolds under discussion. At first sight it might appear hopeless
to find a construction corresponding to the analysis of (2,1)--forms
because of the following example which involves the orbifold of a
3--torus.

Consider the
orbifold $T_1^3/\ZZ_3^2$ where the two actions are defined as
$(z_1,z_4) \longrightarrow  (\a z_1,\a^2 z_4)$, all other coordinates
invariant
and $(z_1,z_7) \longrightarrow (\a z_1,\a^2 z_7) $, all other invariant.
Here $\a$ is the third root of unity.
The resolution of the singular orbifold leads to a Calabi--Yau manifold
with
 84 antigenerations and no generations [15].
This is precisely the mirror flipped spectrum of the exactly solvable
tensor
model $1^9$ of 9 copies of $N=2$ superconformal minimal models at level
$k=1$ [16] which can be described in terms of the
Landau--Ginzburg
potential $W=\sum z_i^3$ which belongs to the configuration
$\IC_{(1,1,1,1,1,1,1,1,1)}[3]$. After imposing the GSO projection by
modding out a $\ZZ_3$ symmetry this Landau--Ginzburg theory leads to the
same spectrum as the $1^9$ theory.

This Landau--Ginzburg theory clearly is a mirror candidate for the
resolved torus orbifold just mentioned
and the question arises whether a manifold corresponding to
this LG potential can be found. Since the theory does not contain modes
corresponding to (1,1)--forms it seems that the manifold cannot be
K\"ahler and hence not projective. Thus it appears that the
7--dimensional manifold $\IP_8[3]$ whose polynomial ring is
identical to the chiral ring
of the LG theory is merely useful as an auxiliary device
in order to describe one sector of the critical LG string vacuum.
Even though there exists a precise identity between the Hodge numbers
in the middle cohomology group of the higher dimensional manifold and the
middle dimension of the cohomology of the Calabi--Yau manifold this is not
the case for the second cohomology group.

It turns out however, that by looking at the manifolds
(\ref{newmfs}) in a
slightly different way it is nevertheless possible to extract the
second cohomology in a canonical manner (even if there is {\it none}).
The way this works is as follows: the manifolds of type
(\ref{newmfs}) will, in general, not be described by smooth spaces but
will have singularities which arise from the projective identification.
The basic idea now is to associate the existence of antigenerations in
a {\it critical} string vacuum with the existence of singularities in
these higher dimensional {\it noncritical} spaces.

Consider again the simple example related to the tensor
model $1^9$. Its LG theory is described by $\IC^*_9[3]$ the naive
compactification of which leads to
\beq
\IP_8[3]\ni \left\{\sum_{i=1}^9 z_i^3=0\right\}.   \nn
\lleq{ex1}
Counting monomials leads to the spectrum of 84 generations found
previously
for the corresponding string vacuum and because this manifold is
smooth {\it no} antigenerations are expected in this model!
Hence there does not exist a Calabi--Yau manifold that describes the
groundstate
\footnote{ It would seem that a  generalization of
   this 7--dimensional smooth manifold is the infinite class of
   models $\IC_{(1,1,1,1,1,1,1,1,1+3q)}[3+q]$, but since the
   manifolds (\ref{newmfs}) are required to be transverse the only
   possibility is $q=0$.}
{}.
A second theory in the space of all LG vacua with no antigenerations is
\beq
(2^6)_{A^6}^{(0,90)} \equiv \IC^*_{(1,1,1,1,1,1,2)}[4] \ni
\left\{\sum_{i=1}^6 z_i^4 + z_7^2 =0\right\}          \nn
\lleq{ex2}
with an obviously smooth manifold $\IP_{(1,1,1,1,1,1,2)}[4] $.

Vacua without antigenerations are rather exceptional however; the generic
groundstate will have both sectors, generations and antigenerations.
The idea described above to derive the antigenerations works for
higher dimensional manifolds corresponding to different types of critical
vacua but in the following we will illustrate it with two types of such
manifolds. A more detailed analysis can be found in [14].

To be concrete consider the
exactly solvable tensor theory $(1\cdot 16^3)_{A_2\otimes E_7^3}$ with
35 generations and 8 antigenerations which
corresponds to a Landau--Ginzburg theory belonging to the configuration
\beq
\IC^*_{(2,3,2,3,2,3,3)}[9]^{(8,35)}
\eeq
and which induces, via projectivization, a 5--dimensional weighted
hypersurface
\beq
\IP_{(2,2,2,3,3,3,3)}[9] \ni
\left\{p=\sum_{i=1}^3 (y_i^3x_i+x_i^3)+x_4^3=0\right\}.
\lleq{ex3}
with orbifold singularities
\bea
\ZZ_3 &:& \IP_3[3] \ni \left\{p_1=\sum_{i=1}^4 x_i^3=0\right\} \nn \\
\ZZ_2 &:& \IP_2.
\eea
The $\ZZ_3$--singular set is a smooth cubic surface which supports
 seven (1,1)--form as can be easily shown. The
$\ZZ_2$--singular set is just the projective plane and therefore
adds one further (1,1)--form.  Hence the singularities induced on the
hypersurface by the singularities of the ambient weighted projective
space
give rise to a total of eight  (1,1)--forms. A simple count leads to the
result that the subring of monomials of charge 1 is of dimension 35.
Thus we have derived the spectrum of the critical theory from the
noncritical manifold (\ref{ex3}).

It is furthermore possible, by using the singularity structure of the
noncritical manifold to actually construct a
 Calabi--Yau manifold of critical dimension directly
from (\ref{ex3}): Recall that the structure of the singularities of
the weighted hypersurface only involved part of the superpotential,
namely the cubic polynomial
$p_1$ which determined the $\ZZ_3$--singular set described by a surface.
The superpotential thus splits naturally into the two parts
\beq
p=p_1 + p_2
\eeq
where $p_2$ is the remaining part of the polynomial. The idea now is to
consider the product $\IP_3[3] \times \IP_2$ where the factors are
determined by
the singular sets of the higher dimensional space and to impose on this
4--dimensional space a constraint described by  the remaining part of
the polynomial which did not
take part in constraining the singularities of this ambient space.
In the
case at hand this leaves a polynomial of bidegree $(3,1)$ and hence
we are lead to a manifold embedded in
\beq
\matrix{\IP_2 \hfill \cr \IP_3\cr}
\left[\matrix{3&0\cr 1&3\cr}\right]
\lleq{ex3mine}
defined by polynomials
\bea
p_1 &=& y_1^3x_1 + y_2^3x_2 +y_3^3x_3 \nn \\
p_2 &=& \sum_{i=1}^4 x_i^3
\eea
which is precisely the manifold constructed in [17], the exactly
solvable
model of which was later found in [18]. Thus we have shown how
to construct the critical Calabi--Yau manifold
from the noncritical manifold (\ref{ex3}).

A class of manifolds of a different type which can be discussed in
this framework rather naturally is defined by
\beq
\IP_{(2k,K-k,2k,K-k,2k_3,2k_4,2k_5)}[2K]
\lleq{niceclass}
where $K=k+k_3+k_4+k_5$.
It has been shown in [5] that
 these higher dimensional spaces lead to manifold embedded in
\beq
\matrix{\IP_1 \hfill \cr \IP_{(k,k,k_3,k_4,k_5)}\cr}
\left[\matrix{2&0\cr k&K\cr}\right],
\lleq{s3class}
spaces which have which admit [19] a $\si$--model
description via the mean field Landau--Ginzburg representation of
ADE minimal tensor models constructed by Gepner [16].

The geometrical picture that emerges from the constructions above
then is the
following: embedded in the higher dimensional manifold is a submanifold
which is fibered, the base and the fibres being determined by the
singular sets of the ambient manifold. The Calabi--Yau manifold itself
is a hypersurface embedded in this fibered  submanifold.

The examples described so far illustrate the simplest situation
that can appear.
In more complicated manifolds the singularity structure will consist
of hypersurfaces whose fibers and/or base themselves are fibered,
leading to
an iterative procedure. The submanifold to be considered will, in
those
cases, be of codimension larger than one and the Calabi--Yau
manifold will be described by a submanifold with codimension
larger than one as well. To illustrate this point consider the 7--fold
\beq
\IP_{(1,1,6,6,2,2,2,2,2)}[8]~\ni~
\left\{\sum_{i=1}^2 \left(x_i^2y_i + y_iz_i +z_i^4\right)
 +z_3^4+z_4^4+z_5^4=0\right\}
\eeq
which leads to the $\ZZ_2$ fibering
$\IP_1\times \IP_{(3,3,1,1,1,1,1)}[4]$ which in turn leads to the
$\ZZ_3$ fibering $\IP_1\times \IP_1 \times \IP_4[4]$. Following the
splits of the potential thus leads to the Calabi--Yau configuration
\beq
\matrix{\IP_1\cr \IP_1 \cr \IP_4\cr}
\left[\matrix{2&0&0\cr
              1&1&0\cr
              0&1&4\cr}\right] \ni
\left\{ \begin{array}{c l}
        p_1=& \sum_{i=1}^2 x_i^2y_i=0 \\
        p_2=& \sum_{i=1}^2 y_iz_i =0 \\
        p_3=& \sum_{j=1}^5 z_i^4 =0
         \end{array}
\right\}
\lleq{codim3}
which is of codimension 3. This example also shows that there are
nontrivial relations between these higher dimensional manifolds.
The way to see this is via the process of splitting and contraction
of Calabi--Yau manifolds introduced in ref. [20].
It can be shown in fact that the Calabi--Yau manifold (\ref{codim3})
is an ineffective split of a Calabi--Yau manifold in the class
(\ref{niceclass}). Thus there also exists a corresponding
relation between the higher dimensional manifolds.

\vskip .3truein
\noindent
{\bf 4. Generalization to Arbitrary Critical Dimensions}

\noindent
Even though the examples discussed in the previous section were all
concerned with 6--dimensional Calabi--Yau manifolds and the way they
are embedded in the new class of spaces, it should be clear that
the ideas presented are not specific to this dimension.
A generalization of an infinite series described in [21]
is furnished by the doubly infinite series of manifolds
\beq
\IP_{(m+1,n-1,m+1,n-1,\dots ,m+1,n-1,m+1,\cdots ,m+1)}[(m+1)n]
\eeq
with $(m+1)$ pairs of coordinates with weights $(m+1,n-1)$  and $(n-m)$
coordinates of weight $(m+1)$, defined by polynomials
\beq
p=\sum_{i=1}^{m+1} (x_i^n+x_iy_i^{m+1}) + \sum_{j=m+2}^{n+1} x_j^n.
\lleq{infser}
According to the considerations above these $(n+1$--dimensional
spaces lead to Calabi--Yau manifolds embedded in
\beq
\matrix{\IP_m\cr \IP_n\cr}\left[\matrix{(m+1)&0\cr 1&n\cr}\right]
\eeq
via the equations
\beq
p_1 = \sum_i y_i^{m+1} x_i,~~~~~p_2 = \sum_{i=1}^{n+1} x_i^n.
\eeq

The simplest example is, of course, the case $n=2$ where the higher
dimensional manifold is a 3--fold described by
\beq
\IP_{(2,1,2,1,2)}[4]~\ni ~
\left\{\sum_{i=1}^2 (z_i^2+z_iy_i^2)+z_3^2=0\right\}
\eeq
with a $\ZZ_2$--singular set isomorphic to the sphere
$\IP_2[2]\sim \IP_1$
which contributes one (1,1)--form, the remaining one being provided
by the
$\IP_1$ defined by the remaining coordinates.
The singularity structure of the 3--fold then relates this space to the
complex torus described by the algebraic curve
\beq
\matrix{\IP_1 \cr  \IP_2}\left[\matrix{2 &0\cr  1 & 2\cr}\right].
\eeq
The Landau--Ginzburg theory corresponding to this
theory derives from an exactly solvable tensor model $(2^2)_{D^2}$
described by two $N=2$ superconformal minimal theories at level $k=2$
equipped with the affine D--invariant.

It is of interest to consider the cohomology groups of the 3--fold
itself.
With the third Chern class $c_3=2h^3$ the Euler number of the
singular space is
\beq
\chi_s = \int c_3 =1
\eeq
and hence the Euler number of the resolved manifold is
\beq
\tchi=1-(2/2)+2\cdot  2 =4.
\eeq
Since the singular set is a sphere its resolution contributes just one
(1,1)--form and hence the second Betti number becomes $b_2=2$.
With  $\tchi=2(1+h^{(1,1)})-2h^{(2,1)}$ it follows that
\beq
h^{(2,1)}=1.
\eeq

The series (\ref{infser}) can be generalized to weighted spaces as is
illustrated by the following example leading to a 4--dimensional critical
manifold:
\beq
p=\sum_{i=1}^3 \left(x_i^3+x_iy_i^3\right) +
  \sum_{j=4}^5 x_i^6
\eeq
 corresponds to the tensor model
$(16^3 \cdot 4)_{E^3 \otimes A^2}$ with central charge $c=12$ and
belongs to the configuration
\beq
\IP_{(6,4,6,4,6,4,3,3)}[18].
\eeq
The critical manifold derived from this 6--fold belongs to
the configuration class
\beq
\matrix{\IP_2 \hfill\cr \IP_{(2,2,2,1,1)}\cr}
\left[\matrix{3&0\cr 2&6\cr}\right]
\eeq
which is indeed a Calabi--Yau deformation class.

A further infinite class [22] of interest consists of the spaces
\bea
\begin{tabular}{l l}
$\IP_{(1,1,....,1,\frac{n+1}{2},\frac{n+1}{2})}[n+1]$,
&$n+1$ even\tabroom \\
$\IP_{(2,2,....,2,n+1,n+1)}[2(n+1)]$, &$n+1$ odd.
\end{tabular}
\eea
of dimension $(n+1)$.
For $(n+1)$ odd the $\ZZ_2$--singular set is a Calabi--Yau manifolds
and the  $\ZZ_{n+1}$--singular set consists of
two points. Hence the higher dimensional space leads to
{\it  two} copies of
the
Calabi--Yau hypersurfaces
\beq
\IP_n[n+1],~~~~ n \in \IN
\eeq
embedded in ordinary projective space.

The simplest case is $n=2$ for which the resolution of the
orbifold singularities of the noncritical 3--fold
\beq
\IP_{(2,2,2,3,3)}[6]
\lleq{counter}
leads to two independent Hodge numbers $h^{(1,1)}=4$, $h^{(2,1)}=2$
and hence the Hodge diamond
      contains {\it twice} the Hodge diamond of the torus, as it
must, according to the geometrical picture described above.
Similarly $\IP_{(2,2,2,2,2,5,5)}[10]$ leads
to two copies of the critical quintic.

The construction is not restricted to the infinite series defined
in (\ref{infser}) or its weighted generalization
 as the next example illustrates.
A five--dimensional critical vacuum of higher codimension
is obtained by considering the Landau--Ginzburg potential
\beq
W= \sum_{j=1}^2 \left(u_i^3 + u_iv_i^2\right) +
    \sum_{i=3}^5 \left(u_i^3 + u_iw_i^3\right)
\eeq
which corresponds to the exactly solvable model
$(16^3\cdot 4^2)_{E_7^3 \otimes D^2}$. The nine--dimensional noncritical
manifold
\beq
\IP_{(3,2,3,2,3,2,3,3,3,3)}[9]
\eeq
leads, via its singularity structure, to the five--dimensional critical
manifold
\beq
\matrix{\IP_1\hfill  \cr \IP_2 \cr \IP_4\cr}
\left[\matrix{2 &0 &0\cr  0 & 3 &0\cr 1 & 1 & 3\cr}\right] .
\eeq
It is crucial that a polynomial was chosen which is not of
Brieskorn--Pham type for the
last four coordinates in the noncritical manifold.

Finally, consider the 9--fold
\beq
\IP_{(5,5,6,6,6,4,4,4,8,8,8)}[16] \ni
\left\{ \sum_{i=1}^2 \left(u_i^2v_i+v_i^2w_i+w_i^2x_i +x_i^2\right)
   + v_3^2w_3+w_3^2x_3 +x_3^2 =0\right\}.
\eeq
The $\ZZ_2$--fibering leads to the split
$\IP_1 \times \IP_{(3,3,3,2,2,2,4,4,4)}$
which in turn leads to a further $\ZZ_2$ split
$\IP_1 \times \IP_2 \times \IP_{(1,1,1,2,2,2)}$
which finally leads to
\beq
\matrix{\IP_1 \cr \IP_2\cr \IP_2 \cr \IP_2\cr}
\left[\matrix{2 &0 &0 &0\cr
              1 &2 &0 &0\cr
              0 &1 &2 &0\cr
              0 &0 &1 &2\cr}\right] \ni
\left\{ \begin{array}{c l}
p_1 =& \sum u_i^2v_i =0 \\
p_2 =& \sum v_i^2w_i =0 \\
p_3 =& \sum w_i^2x_i =0 \\
p_4 =& \sum x_i^2 =0.
         \end{array}
\right\}
\eeq
Thus the 9--fold fibers iteratively and the splits of the polynomial
$p$ are dictated by the fibering.

\vskip .3truein
\noindent
{\bf 5. Cohomology of Noncritical Manifolds}

\noindent
A general relation between the cohomology of the critical vacuum
({\it not} described by a Calabi--Yau manifold in general) and the
cohomology of the higher dimensional space emerges:
\beq
H^{(p,q)}(V_{D_{crit}}) \subset H^{(p+Q-1,q+Q-1)}(M_{D_{crit}+2(Q-1)}).
\eeq
The embedding is nontrivial because the cohomology groups of the
noncritical manifolds are generically
larger than those of the critical vacuum
and hence a projection to the critical spectrum is necessary.
The construction of [5] described above provides a
geometrical framework for such a projection.

One important point regarding the cohomology of the higher dimensional
manifolds of type (\ref{newmfs}) that follows from the considerations of
the previous sections is the fact that $h^{(D_{crit}+Q-1,Q-1)}$
can exceed unity.
Since a Calabi--Yau manifold is defined by the existence of a holomorphic
$(D_{crit},0)$--form one might have expected the noncritical spaces
to be characterized by the existence of a unique
$(D_{crit}+(Q-1),(Q-1))$--form. This is not true as the example
(\ref{counter}) shows. In general the  Hodge--diamond for manifolds
of odd complex dimension leads to a middle Betti number
\beq
b_{D_{crit}+2(Q-1)}=2n+2\sum_{i=1}^{D_{crit}-1}
h^{(D_{crit}+(Q-1)-i,(Q-1)+i)}
\eeq
where $n$ is some integer larger than zero that may exceed unity.
The reason why $n$ may, and sometimes will, be larger than unity
is clear from the geometrical construction reviewed above.

Furthermore, because $h^{(p,p)}(M_{D_{crit}+2(Q-1)})$ is, in general,
larger than zero and
$h^{(Q,Q)}(M_{D_{crit}+2Q}) > h^{(1,1)}(V_{crit})$,
it follows that there exists more than one `remaining' mode,
not accounted for
by the generations and antigenerations of the critical vacuum. This
happens because the noncritical manifolds almost always have
blow--up modes and therefore the cohomology becomes more complicated
than that of the
smooth spaces $\IP_8[3]$ and $\IP_{(1,1,1,1,1,1,2)}[4]$ which
contain only one additional field. In general, then, we have to
expect that not only the dilaton but also other string modes, such
as torsion, will play a role in a possible stringy interpretation.

\vskip .3truein
\noindent
{\bf 6. Conclusion \hfil}

\noindent
It follows from the existence of rigid Calabi--Yau spaces that
mirror symmetry cannot be understood in the framework  of
K\"ahler manifolds with vanishing first Chern class.
To the believer this suggests that beyond the class of such spaces
 there must exist a space of a new type of noncritical manifolds
which contain information about critical vacua, such as the mirrors
of these rigid Calabi--Yau manifolds. Mirrors of spaces with
both sectors, antigeneration and generations, however,
are again of Calabi--Yau type and hence those noncritical
manifolds which correspond to such groundstates should make
contact with Calabi--Yau manifolds in some manner.

What has been  shown in [5] is that the class (\ref{newmfs}) of
higher dimensional K\"ahler
manifolds with positive first Chern class, quantized in a particular way,
generalizes the framework of Calabi--Yau vacua in the desired way: For
particular types of such noncritical manifolds Calabi--Yau manifolds of
critical dimension are embedded algebraically in a fibered submanifold.
For string vacua which cannot be described by
K\"ahler manifolds and which are mirror candidates of rigid Calabi--Yau
manifolds the higher dimensional manifolds still lead to the
spectrum of the critical vacuum and a rationale emerges that explains
why a Calabi--Yau representation is not possible in such theories.
Thus these manifolds of dimension $c/3 +2(Q-1)$
 define an appropriate framework in which to discuss mirror symmetry.

There are a number of important consequences that follow from the
results
of the previous sections. First it should be realized that the relevance
of noncritical manifolds suggests the generalization of a conjecture
regarding
the relation between (2,2) superconformal field theories of central
charge
$c=3D$, $D\in \IN$, with N=1 spacetime supersymmetry on the one hand
and K\"ahler manifolds
of complex dimension $D$ with vanishing first Chern class on the other.
It was suggested
by Gepner that this relation is 1--1. It follows from the results
above that
instead superconformal theories of the above type are in correspondence
with K\"ahler manifolds of dimension $c/3 +2(Q-1)$ with a first Chern
class quantized in multiples of the degree.

A second consequence is that the ideas of section 3 lead, for a large
class of Landau--Ginzburg theories, to a new canonical prescription
for the
construction of the critical manifold, if it exists, directly from
the 2D field theory.

Batyrev [23] has introduced a combinatorical construction of
Calabi--Yau mirrors based on toric geometry. This method appears to
apply only to manifolds defined by one polynomial in a weighted
projective space or products thereof. Because the method used in
[23] is not restricted to Calabi--Yau manifolds
[24] the constructions described in
sections 3 and 4 lead to the possibility of extending Batyrev's
results to Calabi--Yau manifolds of codimension larger
than one by proceeding via noncritical manifolds.

As a final remark it should be emphasized that in this framework
the role played by the dimension of the manifolds parametrizing the
spectrum observed in four dimensions becomes of secondary importance.
This is as it should be,
at least for an effective theory, which tests only matter content and
couplings. It is then, perhaps, not too surprising that via ineffective
splittings manifolds of different
dimension describe one and the same critical vacuum.

It is clear that the emergence in string theory of manifolds with
quantized first Chern class should be understood better. The results
described here are a first step in this direction. They indicate
that these manifolds are not just  auxiliary devices but may be as
physical as  Calabi--Yau manifolds of critical dimension.
In order to probe the structure of these models in more depth it is
important to get further insight into the complete spectrum of
these theories and to compute the Yukawa couplings of the fields.
It is clear from the results presented here that the spectra of the
higher dimensional manifolds contain additional modes beyond those that
are related to the generations and antigenerations of the critical
vacuum
and the question arises what physical interpretation these fields have.

A better grasp on the complete spectrum of these spaces should also
give insight into a different, if not completely independent, approach
toward a deeper understanding of these higher dimensional manifold,
which
is to attempt the construction of consistent $\si$--models defined via
these spaces. Control of the complete spectrum will shed light
on the precise relation between the $\si$--models associated to the
noncritical manifolds and critical $\si$--models.

\vskip .3truein
\noindent
{\bf 7. Acknowledgements \hfil}

\noindent
I'm grateful to the Institute for Theoretical Physics at Santa Barbara
and the Theory Group at the University of Texas at Austin
for hospitality and to Per Berglund, Philip Candelas and Andy Strominger
for discussions.

\vskip .3truein
\noindent
{\bf 8. References \hfil}

\begin{enumerate}
\item P. Candelas, M.Lynker and R.Schimmrigk,
              Nucl.Phys. {\bf B341}(1990)383
\item M.Lynker and R.Schimmrigk, Phys.Lett. {\bf B249}(1990)237
\item B.R.Greene and R.Plesser, Nucl.Phys. {\bf B338}(1990)15
\item B.Zumino, Phys.Lett. {\bf B87}(1979)203
\item R.Schimmrigk, Phys.Rev.Lett. {\bf 70}(1993)3688
\item P.Candelas, E.Derrick and L.Parkes, CERN--TH. 6931/93 preprint
\item C.Vafa, in {\sc Essays in Mirror Symmetry}, ed. S.-T.Yau
\item T.H\"ubsch, unpublished
\item E.Martinec, Phys.Lett. {\bf B217}(1989)431
\item C.Vafa and N.Warner, Phys.Lett. {\bf B218}(1989)51
\item C.Vafa, Mod.Phys.Lett. {\bf A4}(1989)1169
\item P.Candelas, Nucl.Phys. {\bf B298}(1988)458
\item P.Berglund, B.R.Greene and T.H\"ubsch, Mod.Phys.Lett. {\bf A7}(1992)1855
\item R.Schimmrigk, work in progress
\item B.R.Greene, C.Vafa and N.Warner, Nucl.Phys. {\bf B324}(1989)371
\item D.Gepner, Nucl.Phys. {\bf B296}(1988)757
\item R.Schimmrigk, Phys.Lett. {\bf B193}(1987)175
\item D.Gepner, PUPT preprint, December 1987
\item R.Schimmrigk, Phys.Lett. {\bf B229}(1989)227
\item P.Candelas, A.Dale, C.A.L\"utken and R.Schimmrigk,
             Nucl.Phys. {\bf B298}(1988)493
\item R.Schimmrigk, in {\sc Proceedings of the NATO ARW
on Low Dimensional Topology and Quantum Field Theory},
Cambridge, England, 1992 and the {\sc Proceedings of the International
Workshop on String Theory, Quantum Gravity and the Unification of
Fundamental Interactions}, Rome, Italy, 1992
\item R.Schimmrigk, in {\sc Proceedings of the
Texas/PASCOS Meeting}, Berkeley, CA, 1992
\item V.V.Batyrev, University of Essen preprint Nov. 1992
\item V.V.Batyrev, Duke Math.J. {\bf 69}(1993)349
\end{enumerate}
\end{document}